\documentclass[11pt]{article}
\usepackage[DIV10]{typearea}

\usepackage[latin1]{inputenc}
\usepackage[english]{babel}
\usepackage{amsfonts}
\usepackage{amsmath}    
\usepackage{amssymb}
\usepackage{amsthm}
\usepackage[pdftex]{color}
\usepackage{graphicx}

\usepackage[pdftex,colorlinks,pdfpagelabels]{hyperref}
\usepackage[figure,table]{hypcap} 
\hypersetup{
   bookmarksnumbered,
   pdfstartview={FitV},
   pdfpagemode={UseOutlines},
   pdfauthor={Jasper Hasenkamp},
   pdftitle={Dark radiation from the axino solution of the gravitino problem},
   pdfsubject={scientific publication}
   ,citecolor={blue},
   linkcolor={blue},
   urlcolor={blue},
   filecolor={blue}
}
\def\a{\alpha}

\def\g{\gamma}

\def\m{\mu}
\def\n{\nu}

\def\r{\rho}
\def\s{\sigma}
\def\t{\tau}

\def\D{\Delta}

\def\G{\Gamma}

\def\O{\Omega}

\newcommand{\meV}{\text{ meV}}
\newcommand{\eV}{\text{ eV}}
\newcommand{\keV}{\text{ keV}}
\newcommand{\MeV}{\text{ MeV}}
\newcommand{\GeV}{\text{ GeV}}
\newcommand{\TeV}{\text{ TeV}}
\newcommand{\seconds}{\text{ s}}

\newcommand{\years}{\text{ y}}

\newcommand{\axino}{\ensuremath{\widetilde a}}

\newcommand{\order}[1]{\ensuremath{\mathcal{O}\left(#1\right)}}
\newcommand{\mgrav}{m_{3/2}}
\newcommand{\maxino}{m_{\widetilde a}}
\newcommand{\msax}{m_\text{sax}}
\newcommand{\mgluino}{m_{\widetilde g}}
\newcommand{\mlosp}{m_\text{losp}}

\newcommand{\mplanck}{\ensuremath{M_{\text{pl}}}}

\newcommand{\Tr}{T_\text{R}}
\newcommand{\Neff}{N_\text{eff}}

\begin{document}

\date{\mbox{ }}

\title{
{\normalsize
21st of July 2011 \hfill\mbox{}\\}
\vspace{2cm}
\bf Dark radiation from the axino solution\\ of the gravitino problem
\\[8mm]}
%
\author{Jasper Hasenkamp\\[2mm]
{\small\it II.~Institute for Theoretical Physics, University of Hamburg, Hamburg, Germany}\\
{\small\it Jasper.Hasenkamp@desy.de}
}
\maketitle

\thispagestyle{empty}

\vspace{1cm}

\begin{abstract}
Current observations of the cosmic microwave background could confirm an increase in
the radiation energy density after
primordial nucleosynthesis but before photon decoupling.
We show that, if the gravitino problem is solved by a
light axino, dark (decoupled) radiation 
emerges naturally in this period leading
to a new upper bound on the reheating temperature $T_\text{R} \lesssim 10^{11} \GeV$.
In turn, successful thermal leptogenesis might predict such an increase.
The Large Hadron Collider could endorse this opportunity.
At the same time, axion and
axino can naturally form the observed dark matter.
\end{abstract}

\newpage

\section{Introduction}
It is a new opportunity to determine the amount of
radiation in the Universe from observations of the cosmic microwave background (CMB) alone
with precision comparable to that from big bang nucleosynthesis (BBN).
Recent measurements by the Wilkinson Microwave Anisotropy Probe (WMAP)~\cite{Komatsu:2010fb},
 the Atacama Cosmology Telescope (ACT)~\cite{Dunkley:2010ge} and
the South Pole Telescope (SPT)~\cite{Keisler:2011aw} indicate---statistically
not significant---the radiation energy density at the time of photon
decoupling to be higher than inferred from primordial nucleosynthesis
in standard cosmology making use of the Standard Model of particle physics, cf.~\cite{Bowen:2001in,Bashinsky:2003tk}.
This could be taken as another hint for physics beyond the two standard models.
The Planck satellite, which is already taking data,
 could turn the hint into a discovery.

We should search for explanations from particle physics for such an increase in
radiation~\cite{Chang:1996ih,Ichikawa:2007jv}, especially,
because other explanations are missing, if the current mean values are accurate.
Since tailoring an otherwise unmotivated solution to this particular problem
is trivial, the origin of the additional radiation should be motivated from
other problems of cosmology and the Standard Model. Ideally a consistent
cosmology is possible at the same time  and without further ingredients.
In this Letter we propose a natural origin of the late emergence of the additional radiation of exactly this kind
from supersymmetry (SUSY) that itself is well known to be motivated independently.

Supersymmetric theories predict the gravitino as the superpartner of the graviton.
From gravity mediated SUSY breaking, its mass $\mgrav$ is expected to set the scale of the soft SUSY breaking
masses of the superpartner in the
minimal supersymmetric standard model (MSSM) as $m_\text{susy} \sim \order{\mgrav}$.
Since the gravitino interacts only gravitational, its decay~\cite{Falomkin:1984eu} or
 decays into the gravitino~\cite{Moroi:1993mb} generically spoil the success of BBN.
The reheating temperature after inflation $T_\text{R}$ were typically restricted
to low values
 excluding, for example, thermal leptogenesis~\cite{Fukugita:1986hr}, which is a simple and
 elegant explanation for the origin of matter with the requirement $T_\text{R} \gtrsim \order{10^{10} \GeV}$.
In any case, a consistent cosmology has to explain the observed matter composition
of the Universe and at the same time should enable a solution to
the strong CP problem of the Standard Model with the most attractive
solution still found in the Peccei-Quinn (PQ) mechanism~\cite{Peccei:1977hh}.
The spontaneous breakdown of the PQ symmetry gives rise to a goldstone
boson---called axion $a$---accompanied by another scalar---called saxion---
and the axino \axino~carrying the fermionic degrees of freedom of the
axion supermultiplet that has interactions suppressed by the PQ scale
$f_a \gtrsim 6 \times 10^8 \GeV$.

It has been recognised that an axino of mass $\maxino \lesssim \order{\text{keV}}$
~\cite{Tamvakis:1982mw}
as the lightest supersymmetric particle (LSP) with a gravitino next-to-LSP of
mass $\mgrav \sim \order{100 \GeV}$ might provide a natural solution to the
cosmological gravitino problem~\cite{Asaka:2000ew}.
Since any gravitino abundance decays invisibly into axino-axion pairs,
reheating temperatures as high as $T_\text{R} \sim 10^{15} \GeV$ were
claimed to be possible.
Due to their suppressed couplings axion and axino are indeed
decoupled from the thermal bath at such late times when the
gravitino decays.
The axino is light to avoid overproduction.
 At the same time, the lightest ordinary supersymmetric particle (LOSP)
can decay into axino and photon. Altogether, decay problems
and overproduction constraints are circumvented.
Recently, an extensive investigation of the phenomenological viability
of the considered mass hierarchy, $m_\text{susy} > \mgrav > \maxino$,
reports on numerous restrictions of the PQ and MSSM parameter space~\cite{Baer:2010gr}.
In particular, much tighter constraints on the reheating temperature dependent
on the PQ scale and/or arising from the assumed bino LOSP were found.

In this Letter we point out that the reheating temperature is
also constrained from the allowed amount of
relativistic particles decoupled from the thermal bath. Such
``dark radiation`` arises from the gravitino decays, because the
axions and axinos with
$m_a, \, \maxino \ll \mgrav$ are emitted relativistically and thus
contribute to the radiation energy density.
This has not been considered so far.
Since the amount of dark radiation is constrained by different observations,
we find a new upper bound on the reheating temperature, i.e., $\Tr \lesssim 10^{11} \GeV$,
 which is four orders of magnitude tighter than the original one.
This bound is completely independent of the PQ parameter space and its
dependence on the MSSM parameter space is quite limited.
However, any conclusion in this Letter relies on the prerequisite that
the gravitino is lighter than the LOSP and the axino is the LSP.

The coincidence of the required temperature for successful standard thermal leptogenesis
and the new upper bound on the reheating temperature opens
up the opportunity to explain the increase in the  radiation energy density by the decay of the gravitino, that
produces unavoidably dark radiation after BBN.
We show that at the same time axion and axino can naturally form the observed dark matter.

\section{Observational constraints}
 Bounds on the radiation energy density are usually given in terms
 of the effective number of neutrino species $\Neff$ defined by
 \begin{equation}
  \r_\text{rad} = \left( 1 + \Neff \frac{7}{8} \left(\frac{T_\n}{T_\g}\right)^4 \right) \r_\g \, ,
  \label{rhorad}
 \end{equation}
where the radiation energy density $\r_\text{rad}$ is given as a sum of the
energy density in photons $\r_\g=(\pi^2/15) T^4$, the energy density in the neutrinos of
the standard model with $\Neff^\text{SM}=3.046$~\cite{Mangano:2005cc}
 and $T_\n/T_\g = (4/11)^{1/3}$
and any departure from the standard scenario parametrised as $\Neff = \Neff^\text{SM} + \D\Neff$.
Since the radiation energy density affects the expansion rate of the Universe,
$\D \Neff$ can be constrained.
Big bang nucleosynthesis provides a constraint on $\Neff$ consistent with the
standard model~\cite{Simha:2008zj}
\begin{equation}
 \Neff^\text{BBN} = 2.4 \pm 0.4 \quad (68\% \text{ CL}) .
\end{equation}
This constraint seems to be consistent with earlier studies.
A recent study~\cite{Izotov:2010ca} suggests an increased $\Neff$ at BBN
\begin{equation}
 \Neff^\text{BBN} = 3.8 \pm 0.35 \quad (68\% \text{ CL}) \, .
\end{equation}
The large central value arises partly from an adapted short neutron lifetime. Furthermore,
the uncertainty is found to be essentially larger in an independent, simultaneous study~\cite{Aver:2010wq}.
However, these uncertainties open up the possibility of an increased $\Neff$ at all times, i.e., $0<\D\Neff^\text{BBN}=\D\Neff^\text{CMB}$.
This is typically achieved by stable additional degrees of freedom that have once been in thermal equilibrium or
the emergence of dark radiation from particle decays before BBN. Except comments on the second possibility, this will not be discussed
in this Letter.
We assume that $\Neff^\text{BBN}=\Neff^\text{SM}=3.046$ and elaborate on the opportunity that
$\Neff$ increased after the time of primordial nucleosynthesis, i.e., $\D\Neff^\text{BBN}=0 <\D\Neff^\text{CMB}$.

The primordial helium abundance is formed at a cosmic time around $1 \seconds$ and
BBN ends after roughly $20 \text{ min} \simeq 1.2 \times 10^3 \seconds$.
Some $100 \text{ ky}$ later structure formation sets in, when the matter energy density becomes equal to
the radiation energy density at $t_\text{eq} \sim 4 \times 10^{12} \seconds$, and---on
cosmological timescales---shortly after electrons and protons recombine to form neutral hydrogen.
The Universe becomes transparent for photons. We observe the surface of last scattering today
as the cosmic microwave background (CMB).
%
The current constraints from observations of the CMB are much stronger
than previous ones~\cite{Bowen:2001in} but still significantly weaker than those from BBN.
Measurements from the WMAP satellite~\cite{Komatsu:2010fb} using the first and third
acoustic peaks and the ground-based ACT~\cite{Dunkley:2010ge} using
observations of the third through the seventh peaks are complementary, since they
span a broad range of scales. This allows an estimate from the CMB alone~\cite{Dunkley:2010ge}
\begin{equation}
\label{cmbalone}
 (\text{CMB alone - ACT}) \quad \Neff^\text{CMB} = 5.3\pm 1.3 \qquad (68\% \text{ CL}).
\end{equation}
Recently the ground-based SPT published~\cite{Keisler:2011aw}
\begin{equation}
\label{cmbalone2}
 (\text{CMB alone - SPT}) \quad \Neff^\text{CMB} = 3.85\pm 0.62 \qquad (68\% \text{ CL}).
\end{equation}
Combined with measurements of today's Hubble expansion rate $H_0$ and baryon acoustic oscillations (BAO) the current
constraints are~\cite{Dunkley:2010ge}
\begin{equation}
\label{tightneffbound}
 ( \text{ACT }+\text{WMAP } + \text{ BAO } + H_0) \quad \Neff^\text{CMB} = 4.56\pm 0.75 \qquad (68\% \text{ CL}) .
\end{equation}
and~\cite{Keisler:2011aw}
\begin{equation}
\label{tightneffbound2}
 ( \text{SPT }+\text{WMAP } + \text{ BAO } + H_0) \quad \Neff^\text{CMB} = 3.86\pm 0.42 \qquad (68\% \text{ CL}) ,
\end{equation}
respectively.
These limits are consistent with $\Neff^\text{SM}$ at the 2-$\s$ level and
any deviation may well be due to systematic errors. However,
there is a hint of tension and the central values are off the standard value by
 $\D\Neff \sim 0.81 \text{-} 2.3$.\footnote{
We do not take into account Lyman-$\a$ forest data, which may probe $\Neff$ at
even later times and seems to favour even larger central values~\cite{Seljak:2006bg}
disfavouring the standard value at 2-$\s$, but also seems to be afflicted with
large systematic uncertainties.
}
More important, the Planck satellite will significantly increase the precision of CMB observations
to $\D\Neff \simeq 0.26$~\cite{Perotto:2006rj} or even better~\cite{Hamann:2007sb}.
Thus Planck could reveal a difference between $\Neff^\text{SM}(=\Neff^\text{BBN})$ and $\Neff^\text{CMB}$
at the level of 3- to 5-$\s$, if the central values from the current measurements
are accurate;
see also~\cite{Bashinsky:2003tk}. Improvements in the determination of $\Neff^\text{BBN}$ became
crucial.
To have an effect on the effective number of neutrino species measured from the CMB,
it is necessary that the additional radiation is generated before
the observable modes of the CMB have reentered the horizon. This
requirement constrains the maximum lifetime $\t^\text{max} < 1650 \years \simeq 5.2 \times 10^{10} \seconds$~\cite{Fischler:2010xz}.
This time is before the time of matter-radiation equality,
i.e., $\t^\text{max}_\text{CMB} < t_\text{eq}$.

The late emergence of visible radiation like photons is practically excluded.
This would not only spoil the success of BBN, but
also induce a chemical potential for the photons that is
bounded by CMB observations.
\section{Emergence of dark radiation}
The observations discussed in the previous section constrain the emergence of dark radiation.
Not to affect BBN and to have an effect on measurements of $\Neff$ from the CMB, the lifetime of a
decaying matter particle is constrained to lie in the range
\begin{equation}
\label{lifetimerange}
 t^\text{end}_\text{BBN} \sim 1.2 \times 10^3 \seconds < \t < 5.2 \times 10^{10} \seconds \simeq \t^\text{max}_\text{CMB}
\end{equation}
or, equivalently, the cosmic temperature at the particle decay
$
 T^\text{end}_\text{BBN} \sim 33 \keV > T^\text{dec} > 5 \eV \simeq T^\text{min}_\text{CMB}
$.
Since $\t^\text{max}_\text{CMB} < t_\text{eq}$ in standard cosmology the Universe is radiation dominated at these
times, so that $T \propto t^{-1/2} $.

We estimate the gravitino decay width with the following considerations.
In the gravitino center-of-mass frame its two-body decay width is in general given by
$
\G_{3/2}= |\mathcal{M}|^2 |\vec p_1|/(8\pi \mgrav^2)\,  .
$
If both decay products are much lighter than the gravitino, their equal with opposite sign 3-momenta, $|\vec p_1|$=$|\vec p_2|$,
are well approximated by the leading term in $|\vec p_1|=\frac{1}{2}(\mgrav-\order{m_1,m_2})$.
The squared amplitude $|\mathcal{M}|^2$ is on dimensional grounds $\propto \mgrav^4/\mplanck^2$,
where $\mplanck=2.4 \times 10^{18} \GeV$ denotes the reduced Planck mass.
We have to sum over the outgoing axino spins and average over the incoming gravitino spins, which gives a factor
$2/4=1/2$. The rest of the squared amplitude will be a factor
$1/6 \times (1 \pm c' (\maxino/\mgrav)^2 \pm c'' (\maxino/\mgrav)^4 \pm \ldots)$, where at the leading order
the gravitino polarisation sum gives a factor $2/3$. We assume that a detailed analysis renders
the typical additional factor $1/4$. The coefficients $c',c'',\ldots$ are usually $1\leq c',c''\leq 12$.
Altogether, we estimate the gravitino decay width
$
 \G_{3/2} \simeq \mgrav^3/(192 \pi \mplanck^2)
$
with an expected error of the order $\maxino/\mgrav$.
As $\t = \G^{-1}$, we obtain the gravitino lifetime
\begin{equation}
\label{gravitinolifetime}
 \t_{3/2} \simeq 2.3 \times 10^9 \seconds \left(\frac{10^2\GeV}{\mgrav}\right)^3
\end{equation}
or, equivalently,
$
 T_{3/2}^\text{dec} \simeq 24 \eV (\mgrav/(10^2 \GeV))^{3/2}
$.
We see that the gravitino decay is expected to happen within the considered range~\eqref{lifetimerange}.
The mass range for the gravitino to decay within~\eqref{lifetimerange} is $10^4 \GeV \gtrsim \mgrav \gtrsim 35 \GeV$.
Thus most sparticle spectra expected at the Large Hadron Collider (LHC) are allowed
from this perspective and as soon as SUSY is discovered the gravitino next-to-LSP
mass is constrained to a small window.
Since the axino is the LSP, the gravitino next-to-LSP decays into axino and axion.
Other gravitino channels emitting gluons or photons are never
restrictive. Due to an additional PQ vertex suppressed by
the PQ scale, the branching ratio of such processes
is extremely small. We find the emitted energy at least
twelve orders of magnitudes below current limits.

Qualitatively, this holds true even if $R$-parity were broken as long as the breaking is not too large.
$R$-parity violating decays of the gravitino in two standard model particles were still gravitational
decays and additionally suppressed by the $R$-parity violating coupling. Therefore, the viability of
the late emergence of dark radiation in this scenario does not depend on conserved $R$-parity.
We expect the upper bound on $R$-parity violating couplings from leptogenesis to be more constraining.

Since the axion and also the axino are much lighter than the decaying gravitino, both
are emitted with relativistic momenta, $p_{a}\simeq p_{\axino} \simeq m_{3/2}/2$.
Therefore, the energy density in dark radiation $\r_\text{dr}$ consisting of the axion and the axino is
 given by the energy density of the gravitino at its decay. It is
\begin{equation}
 \r_\text{dr}|_{T^\text{dec}_{3/2}} = \r_a|_{T^\text{dec}_{3/2}} +\r_{\axino}|_{T^\text{dec}_{3/2}}
 = \r_{3/2}|_{T^\text{dec}_{3/2}} = \mgrav Y^\text{tp}_{3/2} s(T^\text{dec}_{3/2}) \, ,
\label{rhodr}
\end{equation}
where the entropy density $s= (2\pi^2/45) g_{\ast s} T^3$ could depend on the number of relativistic
degrees of freedom in the thermal bath $g_\ast$ and their corresponding temperatures. However, for the times
considered in this section $g_{\ast s}=3.91$ is fix.
If we would not make use of the sudden-decay approximation, the resulting energy density
would differ by a factor $\sqrt{\pi}/2 \simeq 0.89$ only~\cite{Covi:2001nw}.
The thermally produced gravitino yield $Y^\text{tp}_{3/2}$ is estimated as~\cite{Bolz:2000fu}
\begin{equation}
\label{gravitinoyield}
 Y^\text{tp}_{3/2} \simeq 1.2 \times 10^{-11} \left(\frac{\mgluino}{1 \TeV}\right)^2
 \left(\frac{10^2 \GeV}{\mgrav}\right)^2
 \left(\frac{\Tr}{10^{10}\GeV}\right)\, ,
\end{equation}
where, as in the following, $\mgluino=\mgluino(\m)$ denotes the gluino mass at $\m \sim 90\GeV$.
Since at times after BBN $g_{\ast s}$ is fix, the energy density of dark radiation scales equal to the energy density of radiation
in the thermal bath, even though dark radiation is not coupled to the bath.
Thus it is most easy to calculate $\D\Neff$ at the gravitino decay.

Using~\eqref{rhodr} and~\eqref{rhorad}
the change of the effective number of neutrino species by the
gravitino decay at any time after BBN
is given by
\begin{equation}
\label{ourdeltaneff1}
 \D N_\text{eff} \simeq 0.6
 \left(\frac{10^2 \GeV}{\mgrav}\right)^\frac{5}{2}
 \left(\frac{\mgluino}{1 \TeV}\right)^2
 \left(\frac{T_R}{10^{10} \GeV}\right) \, .
\end{equation}
This is a number of $\order{1}$ and could have been a priori anything. This coincidence for
parameter values motivated by completely disconnected reasons is the key
observation of the present work.
We remind that the gluino is expected to be
among the heaviest sparticles and the gravitino next-to-LSP with $m_\text{susy}\sim \mgrav \sim 10^2\GeV$
is not only motivated from SUSY breaking, but is part of the solution of the gravitino
problem as described in the introduction. This opens up
the opportunity for thermal leptogenesis that requires a reheating temperature
sufficiently larger than $2 \times 10^9 \GeV$. The crucial experimental
input to determine the minimal temperature are future measurements of the light neutrino masses.
\begin{figure}[tb]
 \centering
   \includegraphics[width=\textwidth]{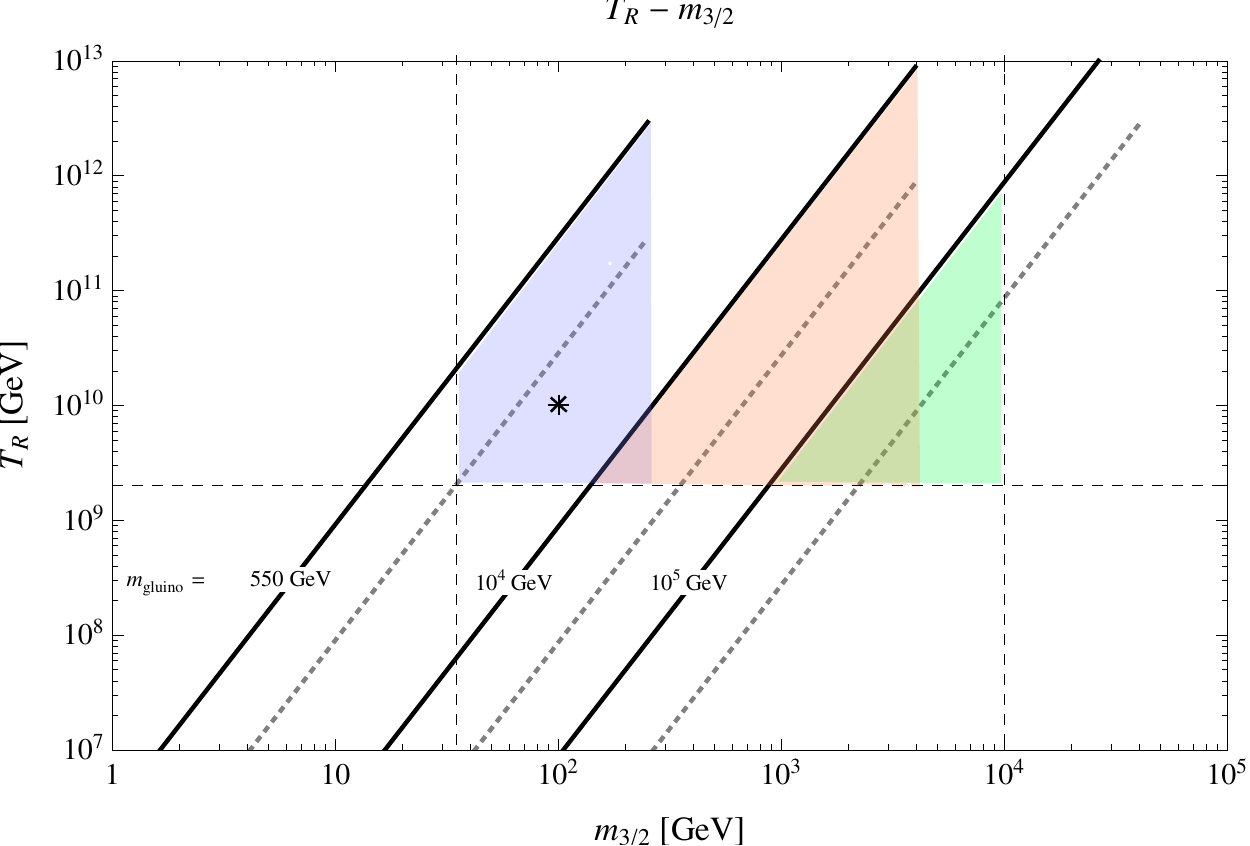}
\caption[$\mgrav$ vs.~$\Tr$]{\small
The solid lines represent upper bounds on the reheating temperature $\Tr$~using~\eqref{ourdeltaneff1} and~\eqref{tightneffbound}as function of the gravitino mass $\mgrav$ for
three different values of the gluino mass $\mgluino= 550 \GeV, 10^4 \GeV$ and $10^5 \GeV$. We restrict $\mgrav < 2 \,\mgluino$.
The shaded regions show the corresponding parameter space consistent with our scenario.
Indicated are bounds from late and early enough gravitino decay~\eqref{lifetimerange}. As discussed
 successful thermal leptogenesis requires $\Tr$ sufficiently larger than $2 \times 10^9 \GeV$.
The dotted lines correspond to values of $\D\Neff=0.52$. This is the possible Planck 2-$\s$ exclusion limit,
if the observed central value coincides with the standard model expectation.
The star corresponds to parameter values as appearing in~\eqref{ourdeltaneff1}.
}
\label{fig:m32-Tr}
 \end{figure}

For the approximations made and fixed sparticle masses as in~\eqref{ourdeltaneff1} we could state that
$\Tr > 9.24 \times 10^{10} \GeV$ is excluded at the 5-$\s$ level referring to
the bound~\eqref{tightneffbound} on the effective number of neutrino species.
For a fixed reheating temperature the gravitino cannot be much lighter, since its mass has the
largest exponent in~\eqref{ourdeltaneff1} and the gluino mass is bounded
from below by experiments. The other way around, for a fixed mass ratio a heavier gravitino does not allow
for much higher reheating temperatures.
Thus we find an upper bound, $T_\text{R} \lesssim 10^{11} \GeV$, four orders of magnitude tighter than in the original
work~\cite{Asaka:2000ew}.
Different upper bounds on the reheating temperature depending on the gravitino mass are depicted in
Fig.~\ref{fig:m32-Tr} for different values of the gluino mass.
The upper bound does not depend at all on the
parameters describing the axion multiplet.
Even if the axino were allowed to be heavy, such that it would not be emitted relativistically,
the emitted radiation energy would reduce by a factor of 2 only. This is due to the axion
that still carries half of the energy.
In this sense, the upper bound does not rely on the very light axino.
A dependence on the MSSM parameters enters
through the gravitino yield~\eqref{gravitinoyield} only.
Since a certain, finite mass gap between gluino and gravitino next-to-LSP
is well-motivated, we find this dependence quite limited.
That $\D\Neff$ in~\eqref{ourdeltaneff1} is of order one opens up another possibility.
An increase in $\Neff$ as discussed in the previous section might be predicted by
the scenario of successful thermal leptogenesis where a light axino solves the gravitino problem,
because the, in general unknown, reheating temperature becomes practically fixed.\footnote{
In higher-dimensional theories the gravitino density may be independent of
$T_\text{R}$ for temperatures required by thermal leptogenesis~\cite{Buchmuller:2003is}.
Larger $T_\text{R}$ were allowed while the predicted increase in $\D\Neff$ were
unchanged.
}
This is especially appealing in a time where current (on-going) experiments
might measure two of the unknowns in~\eqref{ourdeltaneff1}.
As mentioned above the Planck satellite mission might reveal a $\D\Neff>0$
and the LHC as collider experiment
might measure the gaugino masses, in particular the gluino mass.
Furthermore, as soon as SUSY is discovered this sheds light on the mass of a
gravitino  assumed  lighter than the lightest superpartner in the MSSM.
The collider phenomenology is not distinguishable from the axino LSP case
without gravitino, which could itself be mistaken as the neutralino LSP case.
However, in contrast to give an example heavy charged LOSPs may leave the detector.
This would immediately imply that dark matter is not formed by neutralinos.
Interestingly, the reheating temperature in an axino LSP scenario might be probed
at the LHC~\cite{Choi:2007rh,Freitas:2011fx}.
It is important to know at what time the emitted particles become nonrelativistic.
In general, particles become nonrelativistic at the temperature when their
momenta become equal to their mass, i.e.,~$p(T^\text{nr}) \simeq m$.
In the rest frame of a particle decaying into two much lighter particles the decay products
carry a momentum $p (T^\text{dec}) \simeq m/2$, where $m$ denotes the mass of the decaying
particle. After the decay these momenta decrease due to the expansion of the Universe.
In our scenario this yields for the axino
\begin{equation}
p_{\axino}(T) = \frac{\mgrav}{2} \left(\frac{g_{\ast s}(T)}{g_{\ast s}(T^\text{dec}_{3/2})}\right)^\frac{1}{3} \frac{T}{T^\text{dec}_{3/2}} \, .
\end{equation}
Assuming $g_{\ast s}(T^\text{dec}_{3/2})=g_{\ast s}(T^\text{nr})$
we obtain
\begin{equation}
 T^\text{nr}_{\axino} \simeq 5 \times 10^{-4} \meV
 \left(\frac{\maxino}{1 \keV}\right)
 \left(\frac{\mgrav}{10^2 \GeV}\right)^\frac{1}{2} \, .
\end{equation}
Since this temperature is much smaller than the temperature of the
Universe today $T_0 \simeq 0.237 \meV$, axinos of mass $\maxino \sim \order{\text{keV}}$ were surely still
relativistic today and thus contribute as radiation to today's energy budget of the Universe.
For larger axino masses $\maxino \sim 200 \keV$ these axinos
would become nonrelativistic today and for even larger masses $\maxino > 1 \MeV$
these axinos became nonrelativistic in between. They were relativistic at photon decoupling
and nonrelativistic today.
However, the contribution of these axinos as dark matter is
suppressed by the small mass ratio $\maxino/\mgrav$.
Since the axion is much lighter than the axino, as $m_a \lesssim 10 \meV$,
 the emitted axions are still relativistic today for all possible parameter values.

If we require the gravitino to never dominate the energy density of the Universe,
we find a constraint on $T_\text{R}$ similar to the one found above.
This is expected, because the Universe
is dominated by radiation around the gravitino decay and the decay emits
a substantial amount of radiation energy.
However, there is no significant amount of entropy produced in the gravitino decay
even if it dominates at its decay, because the decay products do not
thermalise. Thus cosmological abundances are not diluted
and a period of matter domination after nucleosynthesis and before matter-radiation
equality is not excluded by these considerations.
Instead, a long period of matter domination is excluded by
measurements of the radiation energy density as~\eqref{cmbalone},~\eqref{cmbalone2},~\eqref{tightneffbound} and~\eqref{tightneffbound2}.

The situation seems more involved for the time window between $\t^\text{max}_\text{CMB}\simeq 5.2 \times 10^{10} \seconds$
and $t_\text{eq} \sim 4 \times 10^{12} \seconds$. It seems reasonable that the release of a huge amount
of dark radiation from a dominating matter particle at these times affects the CMB in an observable way
even though maybe only at lower multipoles.
An analysis of this situation is beyond the scope of this Letter and proposed as future research.
\section{LOSP decay and dark matter}
In the considered scenario neither the gravitino nor the LOSP\footnote{
The lightest ordinary superparticle (LOSP) is the
lightest superpartner in the MSSM.
In this sense, gravitino and axino are extraordinary superparticles.
}
can account for the observed dark matter, but the axion
and the axino might be natural candidates.
As their production depends on the Peccei-Quinn scale $f_a$,
we need to consider constraints on it from LOSP decay.\footnote{
In this Letter $f_a=f_\text{PQ}/N$ is called Peccei-Quinn scale with $f_\text{PQ}$ denoting
the scale of PQ symmetry breaking if $N$ denotes the number of different vacua.  For
the KSVZ (DFSZ) model $N=1 \, (6)$.
}
The LOSP has to decay early enough not to spoil the success of BBN.
We remind that the constraints on the LOSP are greatly relaxed compared to the
gravitino LSP case, because it decays much faster, since $f_a m_\text{losp}/(\mplanck \mgrav)^2 \ll 1$.

Since the lightest neutralino is likely one of the lightest
superparticles, we consider the case of the neutralino being the LOSP and thus
its decay into axino and photon.
In general, the decay width depends on the bino component, the phase space, the axion model
and its implementation~\cite{Covi:2001nw}.
For simplicity we assume
$
\G(\widetilde B \rightarrow \axino + \g) = \a_\text{em}^2 m_{\widetilde B}^3/(128 \pi^3 f_a^2)
$
with $\a_\text{em} = 1/128$.
Using $\G = \t^{-1}$
we obtain the upper bound
\begin{equation}
 f_a \lesssim 2 \times 10^{10} \GeV
\left( \frac{m_{\widetilde B}}{100 \GeV} \right)^\frac{3}{2}
\left( \frac{\t^\text{max}_{\widetilde B}}{10^{-2} \seconds} \right)^\frac{1}{2} ,
\end{equation}
where we demand the bino lifetime  $\t_{\widetilde B}$ to be at most $0.01 \seconds$, which is
supposed to be conservative.
 If also the decay into \axino~and Z boson is kinematically unsuppressed,
 the bound is additionally relaxed by a factor of two.
Note that a substantial wino or Higgsino component lowers the relic density
of the neutralino, such that a much later decay becomes allowed with $\t \sim (10^2 \text{--}10^3) \seconds$~\cite{Covi:2009bk}.
Thus a mixed bino-wino or bino-Higgsino state may allow for $f_a$ even larger than $10^{12} \GeV$
already for neutralino masses close to $\mgrav\sim 100 \GeV$.

The DFSZ~\cite{Dine:1981rt} axino couples to Higgsino-Higgs and sfermion-fermion via dimension-4 operators.
Thus Higgsino and stau can be the LOSP.
These channels easily allow for $f_a$ even larger than $10^{12} \GeV$ as soon
as they are kinematically open.
Estimates for any LOSP candidate can be taken from~\cite{Hasenkamp:2011xh}
and references therein, respectively.\footnote{
In~\cite{Hasenkamp:2011xh} the decay $\axino \rightarrow \text{LOSP} +\ldots$ has been considered.
Thus LOSP and axino need to be interchanged in the process. A recent comprehensive study of stau LOSP
constraints is found in~\cite{Freitas:2011fx}.
}
For LOSP masses as large as the upper bound on the gravitino mass, $\mgrav \lesssim 10^4 \GeV$
as following from~\eqref{gravitinolifetime},
Peccei-Quinn scales as large as $10^{13} \GeV$ become allowed already for a bino LOSP
and other LOSP candidates might even allow for $f_a\sim 10^{16}\GeV$.

Axino and axion are produced in the early universe.
The axino is produced in thermal scatterings after inflation, while the axion is dominantly
produced nonthermally from the vacuum misalignment mechanism~\cite{Preskill:1982cy} or possible
topological defects~\cite{Sikivie:1982qv}.
Due to its tiny mass the thermally produced axion density is negligible.

In the KSVZ~\cite{Kim:1979if} model with heavy exotic quarks the relic axino energy density
in units of today's critical energy density $\O_{\axino} = m_{\axino} Y_{\axino} s_0/\r_0$ with
$s_0/\r_0 \simeq 5.4 \times 10^8 \GeV^{-1}$ has been calculated in~\cite{Covi:2001nw,Brandenburg:2004du} to
\begin{equation}
\label{Oaxino}
 \O_{\axino}^\text{ksvz} \sim 0.1
\left( \frac{\maxino}{1 \keV} \right)
\left( \frac{T_\text{R}}{10^{10} \GeV} \right)
\left( \frac{10^{12} \GeV}{f_a} \right)^2 \, .
\end{equation}
This is already close to its density if the axino enters thermal
equilibrium $\O_{\axino}^\text{eq}\simeq 1\times (\maxino/1\keV)$.
A recent study~\cite{Bae:2011jb}
reports that this abundance can be suppressed, if the
KSVZ quarks are lighter than the scale of PQ symmetry breaking. Furthermore,
the axino yield is reported to become independent of the reheating temperature
in the DFSZ model,
\begin{equation}
\O_{\axino}^\text{dfsz} \sim 10^{-6}
\left(\frac{\maxino}{1 \keV}\right)
\left(\frac{10^{12}\GeV}{f_a}\right)^2
(+ \O_{\axino}^\text{freeze-in})
\end{equation}
 for $T_\text{R} \gtrsim 10^4 \GeV$, where we also indicate that
the DFSZ axino is in addition produced via ``freeze-in'' from heavy Higgsino and/or Higgs decays~\cite{Chun:2011zd}.
The yield from freeze-in is again proportional to $f_a^{-2}$ and independent of the reheating temperature,
but dependent on the Higgsino and Higgs masses.
In the following three paragraphs we sketch different example scenarios to account for
the observed dark matter in the considered setting.\footnote{
Of course this list does not contain all possibilities to achieve a consistent cosmology.
}

\vspace{1mm}
{\it I Natural cold axion dark matter:}
The density of cold axions from vacuum misalignment,
\begin{equation}
\O_a^\text{mis} \sim \Theta^2 \left( \frac{N f_a}{10^{12} \GeV} \right)^\frac{7}{6} ,
\end{equation}
is of the order of the observed dark matter density $\O_\text{DM} \simeq 0.21$ for an initial misalignment angle $\Theta$ of
order one and $f_a$ not too far below $10^{12} \GeV$~\cite{Wantz:2009it}.
 As we have seen, constraints from BBN on the decays of different
LOSP candidates often allow for large enough Peccei-Quinn scales with the important exception of a
bino-like LOSP with $m_{\widetilde B} \lesssim 500 \GeV$.
At the same time topological defects do not occur, because the scale
of PQ symmetry restoration is larger than the reheating temperature.
From~\eqref{Oaxino} we see that the axino is required to be light, i.e., $\lesssim \order{\text{keV}}$
to make up only a small fraction of $\O_\text{DM}$. We mention that an admixture of such light
axinos could be favoured from problems of small scale structure formation~\cite{Moore:1999nt}.
The DFSZ axino might be allowed heavier, $\maxino^\text{dfsz} \sim 200 \keV \text{ -- } 1\MeV$,
but this depends on the relative contributions from the different production mechanisms.
Altogether, for $f_a \sim 10^{12} \GeV$ the dark matter is naturally formed by
cold axions and the axino is required to be light.

{\it II Warm axino dark matter:}
For smaller $f_a \sim 10^{10} \GeV$ the generic axion density becomes negligible and
the KSVZ axino is overproduced.\footnote{
For the KSVZ model symmetry restoration would not be problematic here. For the DFSZ model
we need to ensure that the PQ symmetry is not restored  for $T_\text{R} \lesssim 10^{10} \GeV$.
}
 Its mass would be required to be at most of \order{\text{eV}}
to satisfy hot dark matter constraints.
This situation is disfavoured, because
 we would lack a natural dark matter candidate and models with such small masses seem hard to achieve.
The DFSZ axino, however, could constitute warm dark matter with $\maxino > \order{\text{keV}}$ in this case.
Thereby it could even be dominantly produced by freeze-in.
Altogether, for $f_a \sim 10^{10} \GeV$ the dark matter could be formed by
warm DFSZ axinos. Much smaller Peccei-Quinn scales are disfavoured
for both axion models, because the reheating temperature would need to be lowered,
so that standard thermal leptogenesis were excluded.

{\it III Beyond LHC:}
From the above discussion we have seen, that PQ scales as large as $10^{16} \GeV$ with correspondingly
heavier sparticles are not forbidden for many LOSP candidates.
 Surely, $\O_a^\text{mis}$ needs to be suppressed by a
vanishing misalignment angle in this case and such heavy sparticles
are disfavoured in order to stabilise the Higgs mass.
  The KSVZ as well as the DFSZ
axino could form cold dark matter in this scenario with a much smaller
mass gap to the gravitino. In this case, the axino is considerably produced
from LOSP decays~\eqref{Oaxinolospdecay}, such that a LOSP with a
small freeze-out abundance is required, and also the gravitino produces
a substantial amount of cold axinos. Anyway, it is possible---but not arbitrarily---to have
the desired dark radiation in form of axions, a consistent cosmology and
no signal at colliders, because all detectable particles are beyond
the discovery range of the LHC.

\vspace{1mm}
Besides the production from the discussed mechanisms,
 the axino LSP is also produced in decays of the LOSP.
The temperature at which axinos from LOSP decay become nonrelativistic
depends on the LOSP lifetime $\t_\text{losp}$ as
\begin{equation}
 T^\text{nr}_{\axino\text{-losp-dec}} \simeq 24 \meV
 \left(\frac{1 \seconds}{\t_\text{losp}}\right)^\frac{1}{2}
 \left(\frac{\maxino}{1 \keV}\right)
 \left(\frac{10^2 \GeV}{\mlosp}\right)
 \left(\frac{g_\ast(\t_\text{losp})}{10.75}\right)^\frac{1}{12} \, ,
\end{equation}
where $\t_\text{losp}$ is in good approximation independent of the axino mass.
Constraining the LOSP lifetime from BBN to be shorter than
a second, $\t_\text{losp} < 1 \seconds$, we see that it
is likely that axinos from the LOSP decay become nonrelativistic
after CMB and before today.
The contribution of these axinos to today's critical energy density $\O_{\axino}^\text{losp decay}$
is simply given by
\begin{eqnarray}
 \O_{\axino}^\text{losp decay} &=& \frac{\maxino}{\mlosp} \O^\text{fo}_\text{losp} \nonumber \\
 			&=& 10^{-8} \left(\frac{\maxino}{1 \keV}\right)
			\left(\frac{10^2 \GeV}{\mlosp}\right) \O^\text{fo}_\text{losp} \, ,
\label{Oaxinolospdecay}
\end{eqnarray}
where the mass hierarchy between LOSP and axino can usually not be reduced much.
The LOSP energy density from freeze-out prior its decay $\O^\text{fo}_\text{losp}$
is highly model and parameter dependent. For weakly interacting
LOSPs simple estimates tend to $\O^\text{fo}_\text{losp} \sim \O_\text{DM}\sim 0.2$.
However, we see that even for large values $\O^\text{fo}_\text{losp} \sim 10^4 \times \O_\text{DM}$,
as they might occur for a bino LOSP~\cite{Covi:2009bk},
axinos from LOSP decay with $\maxino \ll m_\text{losp}$ give a negligible contribution to $\O$
after they became nonrelativistic.

On the other hand, axinos from LOSP decay contribute to
the radiation energy density,
 as those from gravitino decay, before they become nonrelativistic.
We take into account the scaling of the energy density of dark
radiation $\r_\text{dr} \propto g_{\ast s}^{4/3}(T) T^4$, while we
simplify exploiting $g_\ast(\t_\text{losp})=g_{\ast s}(\t_\text{losp})$
and assuming $g_{\ast s}(\t_\text{losp})>g_{\ast s}(T)=3.91$.
The resulting change of the effective number of neutrino species is
\begin{equation}
  \D\Neff^\text{losp decay} \simeq 9 \times 10^{-6} \, \O^\text{fo}_\text{losp}
  \left(\frac{\t_\text{losp}}{1 \seconds}\right)^\frac{1}{2}
  \left(\frac{r_{\axino}}{0.5}\right)
  \left(\frac{10.75}{g_\ast (\t_\text{losp})}\right)^\frac{1}{12} ,
\end{equation}
where  $r_{\axino}$ denotes the fraction of emitted energy that is carried
by the axino. If the standard model particle  emitted together with the axino
is also very light or massless as the photon, $r_{\axino}=0.5$ is a good approximation.
With the same reasoning as after~\eqref{Oaxinolospdecay} we see that $\D\Neff$ from LOSP
decay is negligible. This is to some extent self-consistent as larger $\t_\text{losp}$
becomes allowed only for smaller $\O^\text{fo}_\text{losp}$.
\paragraph{Saxion}
Up to now we have not discussed cosmological constraints from and on the saxion.
Usually the saxion leads to cosmological problems from too late decay
 as the other particles with suppressed  couplings.
 However, since  in the considered scenario the saxion is allowed to produce LOSPs at any time before
BBN\footnote{
This is the same in the case of broken $R$-parity which is discussed in~\cite{Hasenkamp:2011xh}.
}, constraints from and on the thermally produced saxion are absent or
moderate in all considered dark matter scenarios, with the
exception of the requirement of a small initial oscillation
amplitude $\phi_\text{sax}^\text{i}\lesssim f_a$ after inflation~\cite{Hasenkamp:2011xh}.

In the absence of the LOSP decay problem, the lower bound on the
saxion mass from early enough decay becomes~\cite{Hasenkamp:2011xh}
\begin{equation}
\label{msaxLospdec}
\msax \gtrsim 108\GeV \left(\frac{f_a}{10^{12}\GeV}\right)^\frac{2}{3} ,
\end{equation}
where we conservatively neglect the saxion decay into an axion pair.
The bound~\eqref{msaxLospdec} is automatically fulfilled for a saxion with $m_\text{sax}\sim m_\text{susy}$ as expected,
when $m_\text{susy}$ denotes the soft SUSY breaking masses. Note that in the considered
case of $f_a\sim 10^{16}\GeV$ we also consider large values of $m_{susy} \gtrsim 10^5 \GeV$.
Thus the bound is fulfilled in this case as well.

The saxion decay into two axions, $\phi_\text{sax}\rightarrow a+a$,
is constraint from the requirement not to produce too much dark radiation
before BBN.
In contrast to the late decaying gravitino as considered in this work,
the saxion has to decay before BBN even if the
production of axion pairs is the dominant decay channel. This is due
to its nonnegligible branching ratio into a gluon pair\footnote{
Considering a saxion with $m_\text{sax}\lesssim 40 \MeV$, produced from coherent oscillations and $\Tr \lesssim 10^6 \GeV$
the saxion could actually lead to a $\D\Neff$ of \order{1} after BBN~\cite{Ichikawa:2007jv}.
}.
However, it leads to an increase in $\Neff$ before BBN, such that
$0<\D\Neff^\text{BBN}=\D\Neff^\text{CMB}$.
First, we consider the constraint on the saxion-axion-axion self-coupling $x$
by requiring that such an increase should be smaller than one.
Then the constraint on the branching ratio of the saxion into two axions $B_{aa}$
reads~\cite{Hasenkamp:2011xh}
\begin{equation}
\label{boundonbaa}
 B_{aa} \lesssim 0.4 \left(1+ 50 \pi^2  x^2 \right)^\frac{1}{2}
  \left(\frac{10^{10} \GeV}{f_a}\right) \left(\frac{\msax}{10^2 \GeV}\right)^\frac{1}{2}
 \left(\frac{Y_\text{sax}^\text{eq}}{Y_\text{sax}}\right) \left(\frac{g_\ast(T_\text{sax}^\text{dec})}{10.75}\right)^\frac{1}{12} \, ,
\end{equation}
where we have approximated the decay width of the
saxion as
$\G_\text{sax} \simeq \G_\text{sax}^{gg} + \G_\text{sax}^{aa}$.
Thus our conclusion should hold qualitatively for any axion model.
Then the branching ratio reduces to
\begin{equation}
 B_{aa} \simeq \frac{\G_\text{sax}^{aa}}{\G_\text{sax}^{aa}+\G_\text{sax}^{gg}} = \frac{x^2}{x^2+2 \a_s^2/\pi^2} \, .
\end{equation}
Since in our scenario the reheating temperature
is supposed to be rather large, for $f_a\sim 10^{10} \GeV$ the saxion enters thermal equilibrium after inflation.
In this case the saxion yield $Y_\text{sax}$ cannot be smaller than the
equilibrium value $Y_\text{sax}^\text{eq} \simeq 1.21 \times 10^{-3}$.
The appearing saxion mass $\msax=10^2\GeV$ in~\eqref{boundonbaa} is chosen to
show the worst situation in the considered scenario.
Like a smaller $f_a$ or a larger $\msax$,
also a larger $x$ leads to
an earlier decay, which corresponds to a smaller $\O_\text{sax}$
at its decay. Thus there is a self-curing effect for large $x$~\cite{Hasenkamp:2011xh}.
 The bound~\eqref{boundonbaa} represents an implicit equation for the
 self-coupling $x$. Evaluating it for $x$ with parameters fixed as appearing in~\eqref{boundonbaa}
  it turns out that there is, indeed,
 no constraint on $x$, because branching ratios are by definition smaller than one.

For larger $f_a\sim 10^{12}\GeV$ and fixed $\Tr$ the saxion does no longer
enter thermal equilibrium. If we assume its yield equals the axino
yield as in~\eqref{Oaxino}, we can insert
\begin{equation}
\label{Ysax}
Y_\text{sax} \simeq 2 \times 10^{-5} \left(\frac{\Tr}{10^{10}\GeV}\right)\left(\frac{10^{12}\GeV}{f_a}\right)^2
\end{equation}
into~\eqref{boundonbaa}. Evaluating again for $x$ we find,
that any constraint on $x$ disappears in the same sense as
before, if $m_\text{sax}\gtrsim 290\GeV$.
Considering $f_a\sim  10^{16}\GeV$ and larger $\msax$
we see from inserting~\eqref{Ysax} into~\eqref{boundonbaa}, that
both parameters, if raised like that, lead to a
vanishing bound on $x$.

Even though $\D\Neff=1$ from saxion decay is consistent with all existing measurements,
for deviations of the effective number of neutrino species $\D\Neff^\text{CMB}< 1.78$ it
would be an unfortunate case considering the discovery potential of Planck. A 3-$\s$ detection
would become impossible.
From the discussion above we see that
$\D\Neff$ from saxion decay can well
 be small for natural parameter values. For $\msax>900 \GeV$ it is expected small in any case and especially for  $x\ll 1$ and $x>1$.
How to achieve a desired value of it, is described in~\cite{Hasenkamp:2011xh}.
One can consider cases with a significant deviation from the
standard model expectation during BBN and an even larger deviation
at photon decoupling, i.e., $0< \D\Neff^\text{BBN} < \D\Neff^\text{CMB}$.
We do not comment further on this possibility.
Note that if the DFSZ saxion can decay into a Higgs pair, this decay might
become stronger than the gluon decay channel.
This reduces the amount of emitted axions.

Altogether, we see that we have to consider the saxion decay into two
axions, especially for the considered scenario. However, depending on the
actually measured value from Planck this leads to acceptable or mild
constraints on the saxion mass. In general, the emergence of dark radiation
before BBN would be constrained by a better determination of the effective
number of neutrino species during BBN.

Note that bounds on the initial oscillation amplitude are not affected
by the absence of the NLSP problem. Practically, we assume a mechanism
that sets the initial amplitude $\phi_\text{sax}^\text{i}\lesssim f_a$.
\section{Conclusions and outlook}
The key observation of the present work is:
The increase in the effective number of neutrino species in~\eqref{ourdeltaneff1}
by the invisible gravitino decay after BBN but before photon decoupling
is of \order{1} for expected (natural) masses and a reheating temperature motivated
by successful standard thermal leptogenesis, which is a completely disconnected notion.
Thereby the presupposed mass hierarchy, $m_\text{susy} > \mgrav > \maxino$,
was motivated to solve the cosmological gravitino problem
by--- in some sense---the cosmological axino problem.
Indeed, we have found a new upper bound on the reheating temperature~$T_\text{R} \lesssim 10^{11} \GeV$
in this scenario.
Since the decay is gravitational, it occurs generically late and
 the bound does not at all depend on
the Peccei-Quinn parameter space. The only requirement on the MSSM
parameter space stems from thermal gravitino production. 
It is a large enough but finite mass gap between
gluino and gravitino next-to-LSP.

Since neither LOSP nor gravitino are dark matter candidates in the considered
scenario, it is not granted that there are natural dark matter candidates at all.
We have found that the axion may naturally form the observed amount of cold dark matter
without any conflict with BBN from late LOSP decay.
To this end, we commented on the BBN constraints from different LOSP candidates
extending previous analyses.
Estimates can be made for any MSSM sparticle using~\cite{Hasenkamp:2011xh}.
We mentioned that an admixture of thermally produced, light axinos may be favoured from small scale problems
of structure formation.
It seems that the DFSZ axino can form warm dark matter for smaller $f_a\sim 10^{10} \GeV$.
The identification of consistent cosmologies
 and their corresponding parameter space with the desired increase in radiation
calls for more comprehensive studies.

We have checked that the scenario is save against other gravitino decay channels and radiation
from LOSP decay. Constraints from and on the saxion are absent or moderate (with
the exception of the bound on its initial oscillation amplitude $\phi_\text{sax}^\text{i}\lesssim f_a$), because it is
allowed to produce LOSPs at any time before BBN as in~\cite{Hasenkamp:2011xh}.
The most severe lower bound on the saxion mass can arise from its decay into an axion pair.
We find these bounds in the worst scenario still acceptable.
The collider phenomenology is not distinguishable from the axino LSP case without gravitino.

Even though we have also shown that the LHC cannot rule out the scenario,
we want to point out its surprisingly high testability considering that
gravitino and axino are generally elusive particles and
the reheating temperature is not an experimentally
accessible quantity.
Combining the new opportunity to
measure the effective number of neutrino species from the CMB alone and
the potential discovery of supersymmetry at the LHC, two of the unknowns in~\eqref{ourdeltaneff1}
can be determined and within the scenario the gravitino mass would be constrained to a small
window as, therefore, the reheating temperature as well.
The viability of thermal leptogenesis in this scenario were tested and
the cosmological gravitino problem would turn out as
a fortune when the Planck mission indeed discovers an increased
radiation energy density.
In turn, that discovery would provide indirect
evidence for this particular scenario.
\subsection*{Acknowledgements}
I am deeply grateful to J\"{o}rn Kersten for his support and encouragement.
This work was supported by the German Science Foundation (DFG) via the
Junior Research Group ``SUSY Phenomenology'' within the Collaborative
Research Centre 676 ``Particles, Strings and the Early Universe''.


\addcontentsline{toc}{chapter}{References}
\begin{thebibliography}{99}
\bibitem{Komatsu:2010fb}
  E.~Komatsu {\it et al.} [ WMAP Collaboration ],
  Astrophys.\ J.\ Suppl.\  {\bf 192}, (2011) 18.
  [arXiv:1001.4538 [astro-ph.CO]].

\bibitem{Dunkley:2010ge}
  J.~Dunkley {\it et al.},
  ``The Atacama Cosmology Telescope: Cosmological Parameters from the 2008
  Power Spectra,''
  arXiv:1009.0866 [astro-ph.CO].

\bibitem{Keisler:2011aw}
  R.~Keisler {\it et al.},
  arXiv:1105.3182 [astro-ph.CO].

\bibitem{Bowen:2001in}
  R.~Bowen, S.~H.~Hansen, A.~Melchiorri, J.~Silk and R.~Trotta,
  ``The Impact of an extra background of relativistic particles on the
  cosmological parameters derived from microwave background anisotropies,''
  Mon.\ Not.\ Roy.\ Astron.\ Soc.\  {\bf 334} (2002) 760
  [arXiv:astro-ph/0110636].

\bibitem{Bashinsky:2003tk}
  S.~Bashinsky and U.~Seljak,
  ``Neutrino perturbations in CMB anisotropy and matter clustering,''
  Phys.\ Rev.\  D {\bf 69} (2004) 083002
  [arXiv:astro-ph/0310198];
  J.~Hamann, S.~Hannestad, G.~G.~Raffelt and Y.~Y.~Y.~Wong,
  ``Observational bounds on the cosmic radiation density,''
  JCAP {\bf 0708} (2007) 021
  [arXiv:0705.0440 [astro-ph]];
  A.~X.~Gonzalez-Morales, R.~Poltis, B.~D.~Sherwin and L.~Verde,
  ``Are priors responsible for cosmology favoring additional neutrino
  species?,''
  arXiv:1106.5052 [astro-ph.CO].

\bibitem{Chang:1996ih}
  S.~Chang and H.~B.~Kim,
  ``A Dark matter solution from the supersymmetric axion model,''
  Phys.\ Rev.\ Lett.\  {\bf 77} (1996) 591
  [arXiv:hep-ph/9604222];
  E.~Calabrese, D.~Huterer, E.~V.~Linder, A.~Melchiorri and L.~Pagano,
  ``Limits on Dark Radiation, Early Dark Energy, and Relativistic Degrees of
  Freedom,''
  Phys.\ Rev.\  D {\bf 83}, 123504 (2011)
  [arXiv:1103.4132 [astro-ph.CO]].

\bibitem{Ichikawa:2007jv}
  K.~Ichikawa, M.~Kawasaki, K.~Nakayama, M.~Senami and F.~Takahashi,
  ``Increasing effective number of neutrinos by decaying particles,''
  JCAP {\bf 0705} (2007) 008
  [arXiv:hep-ph/0703034].

\bibitem{Falomkin:1984eu}
  I.~V.~Falomkin, G.~B.~Pontecorvo, M.~G.~Sapozhnikov, M.~Y.~Khlopov, F.~Balestra and G.~Piragino,
``Low-energy anti-p He-4 annihilation and problems of the modern cosmology, GUT and SUSY models,''
  Nuovo Cim.\  A {\bf 79} (1984) 193
  [Yad.\ Fiz.\  {\bf 39} (1984) 990];
  J.~R.~Ellis, D.~V.~Nanopoulos and S.~Sarkar,
  ``The Cosmology of Decaying Gravitinos,''
  Nucl.\ Phys.\  B {\bf 259} (1985) 175.

\bibitem{Moroi:1993mb}
  T.~Moroi, H.~Murayama and M.~Yamaguchi,
  ``Cosmological constraints on the light stable gravitino,''
  Phys.\ Lett.\  B {\bf 303} (1993) 289.

\bibitem{Fukugita:1986hr}
  M.~Fukugita and T.~Yanagida,
  ``Baryogenesis Without Grand Unification,''
  Phys.\ Lett.\  B {\bf 174} (1986) 45.


\bibitem{Peccei:1977hh}
  R.~D.~Peccei and H.~R.~Quinn,
  ``CP Conservation in the Presence of Instantons,''
  Phys.\ Rev.\ Lett.\  {\bf 38} (1977) 1440;
  R.~D.~Peccei and H.~R.~Quinn,
  ``Constraints Imposed by CP Conservation in the Presence of Instantons,''
  Phys.\ Rev.\  D {\bf 16} (1977) 1791.

\bibitem{Tamvakis:1982mw}
  K.~Tamvakis and D.~Wyler,
``Broken global symmetries in supersymmetric theories,''
  Phys.\ Lett.\  B {\bf 112} (1982) 451;
  J.~E.~Kim,
``A common scale for the invisible axion, local SUSY GUTs and saxino decay,''
  Phys.\ Lett.\  B {\bf 136} (1984) 378;
  J.~F.~Nieves,
``Spontaneous breaking of global symmetries in supersymmetric theories,''
  Phys.\ Rev.\  D {\bf 33} (1986) 1762;
  T.~Goto and M.~Yamaguchi,
  ``Is axino dark matter possible in supergravity?,''
  Phys.\ Lett.\  B {\bf 276} (1992) 103;
  E.~J.~Chun, J.~E.~Kim and H.~P.~Nilles,
  ``Axino mass,''
  Phys.\ Lett.\  B {\bf 287}, 123 (1992)
  [arXiv:hep-ph/9205229].

\bibitem{Asaka:2000ew}
  T.~Asaka and T.~Yanagida,
  ``Solving the gravitino problem by axino,''
  Phys.\ Lett.\  B {\bf 494}, 297 (2000)
  [arXiv:hep-ph/0006211].

\bibitem{Baer:2010gr}
  H.~Baer, S.~Kraml, A.~Lessa and S.~Sekmen,
  ``Thermal leptogenesis and the gravitino problem in the Asaka-Yanagida
  axion/axino dark matter scenario,''
  JCAP {\bf 1104}, 039 (2011)
  [arXiv:1012.3769 [hep-ph]]; for stau LOSP see also~\cite{Freitas:2011fx}.

\bibitem{Freitas:2011fx}
  A.~Freitas, F.~D.~Steffen, N.~Tajuddin, D.~Wyler,
  ``Axinos in Cosmology and at Colliders,''
  JHEP {\bf 1106}, 036 (2011)
  [arXiv:1105.1113 [hep-ph]].


\bibitem{Mangano:2005cc}
  G.~Mangano, G.~Miele, S.~Pastor, T.~Pinto, O.~Pisanti and P.~D.~Serpico,
  ``Relic neutrino decoupling including flavor oscillations,''
  Nucl.\ Phys.\  B {\bf 729}, 221 (2005)
  [arXiv:hep-ph/0506164].

\bibitem{Simha:2008zj}
  V.~Simha and G.~Steigman,
  ``Constraining The Early-Universe Baryon Density And Expansion Rate,''
  JCAP {\bf 0806}, 016 (2008)
  [arXiv:0803.3465 [astro-ph]].

\bibitem{Izotov:2010ca}
  Y.~I.~Izotov and T.~X.~Thuan,
  ``The primordial abundance of 4He: evidence for non-standard big bang
  nucleosynthesis,''
  Astrophys.\ J.\  {\bf 710}, L67 (2010)
  [arXiv:1001.4440 [astro-ph.CO]].
\bibitem{Aver:2010wq}
  E.~Aver, K.~A.~Olive and E.~D.~Skillman,
  ``A New Approach to Systematic Uncertainties and Self-Consistency in Helium
  Abundance Determinations,''
  JCAP {\bf 1005}, 003 (2010)
  [arXiv:1001.5218 [astro-ph.CO]].


\bibitem{Seljak:2006bg}
  U.~Seljak, A.~Slosar and P.~McDonald,
  ``Cosmological parameters from combining the Lyman-alpha forest with CMB,
  galaxy clustering and SN constraints,''
  JCAP {\bf 0610}, 014 (2006)
  [arXiv:astro-ph/0604335].

\bibitem{Perotto:2006rj}
  L.~Perotto, J.~Lesgourgues, S.~Hannestad, H.~Tu and Y.~Y.~Y.~Wong,
  ``Probing cosmological parameters with the CMB: Forecasts from full Monte
  Carlo simulations,''
  JCAP {\bf 0610}, 013 (2006)
  [arXiv:astro-ph/0606227].


\bibitem{Hamann:2007sb}
  J.~Hamann, J.~Lesgourgues and G.~Mangano,
  ``Using BBN in cosmological parameter extraction from CMB: A Forecast for
  PLANCK,''
  JCAP {\bf 0803}, 004 (2008)
  [arXiv:0712.2826 [astro-ph]].

\bibitem{Fischler:2010xz}
  W.~Fischler and J.~Meyers,
  ``Dark Radiation Emerging After Big Bang Nucleosynthesis?,''
  Phys.\ Rev.\  D {\bf 83}, 063520 (2011)
  [arXiv:1011.3501 [astro-ph.CO]].

\bibitem{Covi:2001nw}
  L.~Covi, H.~B.~Kim, J.~E.~Kim and L.~Roszkowski,
  ``Axinos as dark matter,''
  JHEP {\bf 0105}, 033 (2001)
  [arXiv:hep-ph/0101009].

\bibitem{Bolz:2000fu}
  M.~Bolz, A.~Brandenburg and W.~Buchmuller,
  ``Thermal production of gravitinos,''
  Nucl.\ Phys.\  B {\bf 606}, 518 (2001)
  [Erratum-ibid.\  B {\bf 790}, 336 (2008)]
  [arXiv:hep-ph/0012052];
  J.~Pradler and F.~D.~Steffen,
  ``Thermal gravitino production and collider tests of leptogenesis,''
  Phys.\ Rev.\  D {\bf 75}, 023509 (2007)
  [arXiv:hep-ph/0608344].

\bibitem{Buchmuller:2003is}
  W.~Buchmuller, K.~Hamaguchi and M.~Ratz,
  ``Gauge couplings at high temperature and the relic gravitino abundance,''
  Phys.\ Lett.\  B {\bf 574}, 156 (2003)
  [arXiv:hep-ph/0307181].

\bibitem{Choi:2007rh}
  K.~Y.~Choi, L.~Roszkowski and R.~Ruiz de Austri,
  JHEP {\bf 0804}, 016 (2008)
  [arXiv:0710.3349 [hep-ph]].


\bibitem{Covi:2009bk}
  L.~Covi, J.~Hasenkamp, S.~Pokorski and J.~Roberts,
  ``Gravitino Dark Matter and general neutralino NLSP,''
  JHEP {\bf 0911} (2009) 003
  [arXiv:0908.3399 [hep-ph]].


\bibitem{Hasenkamp:2011xh}
  J.~Hasenkamp and J.~Kersten,
 ``Dark and visible matter with broken R-parity and the axion multiplet,''
  Phys.\ Lett.\  B {\bf 701} (2011) 660
  [arXiv:1103.6193 [hep-ph]].

\bibitem{Preskill:1982cy}
  J.~Preskill, M.~B.~Wise and F.~Wilczek,
  ``Cosmology of the Invisible Axion,''
  Phys.\ Lett.\  B {\bf 120}, 127 (1983);
  L.~F.~Abbott and P.~Sikivie,
  ``A Cosmological Bound on the Invisible Axion,''
  Phys.\ Lett.\  B {\bf 120}, 133 (1983);
  M.~Dine and W.~Fischler,
  ``The Not So Harmless Axion,''
  Phys.\ Lett.\  B {\bf 120}, 137 (1983).

\bibitem{Sikivie:1982qv}
  P.~Sikivie,
  ``Of Axions, Domain Walls and the Early Universe,''
  Phys.\ Rev.\ Lett.\  {\bf 48}, 1156 (1982);
  R.~L.~Davis,
  ``Cosmic Axions from Cosmic Strings,''
  Phys.\ Lett.\  B {\bf 180}, 225 (1986).

\bibitem{Kim:1979if}
  J.~E.~Kim,
  Phys.\ Rev.\ Lett.\  {\bf 43}, 103 (1979);
  M.~A.~Shifman, A.~I.~Vainshtein and V.~I.~Zakharov,
  Nucl.\ Phys.\  B {\bf 166}, 493 (1980).




\bibitem{Brandenburg:2004du}
  A.~Brandenburg, F.~D.~Steffen,
  ``Axino dark matter from thermal production,''
  JCAP {\bf 0408}, 008 (2004).
  [hep-ph/0405158];
  A.~Strumia,
  ``Thermal production of axino Dark Matter,''
  JHEP {\bf 1006}, 036 (2010).
  [arXiv:1003.5847 [hep-ph]].


\bibitem{Bae:2011jb}
  K.~J.~Bae, K.~Choi, S.~H.~Im,
  ``Effective interactions of axion supermultiplet and thermal production of axino dark matter,''
  JHEP {\bf 1108 } (2011)  065.
  [arXiv:1106.2452 [hep-ph]];
after completion of the aparent work following publication appeared reporting different suppressions and a
a strong dependence on the IR regulator used in the axino thermal production computation:
  K.~-Y.~Choi, L.~Covi, J.~E.~Kim, L.~Roszkowski,
  ``Axino Cold Dark Matter Revisited,''
  [arXiv:1108.2282 [hep-ph]].

\bibitem{Dine:1981rt}
  M.~Dine, W.~Fischler and M.~Srednicki,
  Phys.\ Lett.\  B {\bf 104}, 199 (1981);
  A.~R.~Zhitnitsky,
  Sov.\ J.\ Nucl.\ Phys.\  {\bf 31}, 260 (1980)
  [Yad.\ Fiz.\  {\bf 31}, 497 (1980)].



\bibitem{Chun:2011zd}
  E.~J.~Chun,
  ``Dark matter in the Kim-Nilles mechanism,''
  arXiv:1104.2219 [hep-ph].

\bibitem{Wantz:2009it}
  O.~Wantz and E.~P.~S.~Shellard,
  ``Axion Cosmology Revisited,''
  Phys.\ Rev.\  D {\bf 82}, 123508 (2010)
  [arXiv:0910.1066 [astro-ph.CO]].

\bibitem{Moore:1999nt}
  B.~Moore, S.~Ghigna, F.~Governato, G.~Lake, T.~R.~Quinn, J.~Stadel and P.~Tozzi,
  ``Dark matter substructure within galactic halos,''
  Astrophys.\ J.\  {\bf 524}, L19 (1999);
  E.~Papastergis, A.~M.~Martin, R.~Giovanelli and M.~P.~Haynes,
  ``The velocity width function of galaxies from the 40
  shedding light on the CDM overabundance problem,''
  arXiv:1106.0710 [astro-ph.CO].

\end{thebibliography}

\phantomsection 

\end{document}